\documentclass[twocolumn,showkeys,preprintnumbers,amsmath,amssymb]{revtex4}
\usepackage{graphicx}
\usepackage{bm}
\usepackage{graphicx}

\usepackage{psfrag}  
\usepackage{amsmath,amsthm}
\usepackage{amsfonts}
\usepackage{amssymb}
\usepackage{bm}

\begin{document}


\title{Percolation of Arbitrary Uncorrelated Nested Subgraphs}  
\author{Bernat  Corominas-Murtra$^1$}

\affiliation{$^1$ ICREA-Complex Systems  Lab, Universitat Pompeu Fabra
  (Parc de Recerca Biom\`edica de Barcelona). Dr Aiguader 88, 08003 Barcelona, Spain}

\begin{abstract}
The study of percolation in so-called {\em nested subgraphs}
  implies  a generalization of  the concept  of percolation  since the
  results are not linked to specific graph process.  Here the behavior
  of such  graphs at  criticallity is studied  for the case  where the
  nesting   operation   is   performed   in   an   uncorrelated   way.
  Specifically, I provide an  analyitic derivation for the percolation
  inequality  showing  that  the  cluster size  distribution  under  a
  generalized process of uncorrelated nesting at criticality follows a
  power law  with universal  exponent $\gamma=3/2$.  The  relevance of
  the result comes from the  wide variety of processes responsible for
  the emergence of  the giant component that fall  within the category
  of  nesting  operations,  whose   outcome  is  a  family  of  nested
  subgraphs.
  \end{abstract}

\maketitle

\section{Introduction}

The behavior of complex networks  under operations of node removal has
attracted the attention of researchers  as one of the main problems in
modern  physics
\cite{Barabasi:2000},\cite{Cohen:2000}\cite{Dorogovtsev:2002},\cite{Callaway},\cite{Random}.
The  general studied  properties are  those based  on either  the {\em
  resilience} of  the net  under some kind  of node (or  link) removal
process  \cite{Barabasi:2000},\cite{Cohen:2000},\cite{Callaway} or its
behavior                         at                        criticality
\cite{Callaway},\cite{Random},\cite{Cohen:2002},   which  is,  roughly
speaking,  what happens  at the  point  where the  process of  removal
reaches    the   objective    of   breaking    completely    the   net
\cite{Barabasi:2000},  \cite{Cohen:2002},  or  more specifically,  the
     {\em Giant Connected  Component} -hereafter, $GCC$ \cite{Random}.
     The  study of  real  systems from  the  view point  of the  above
     mentioned concepts implied a breakthrough in the understanding of
     internet   fragility    \cite{Cohen:2000},   ecological   systems
     \cite{Sole:2001}  or  disease  spreading  \cite{Pastor-Satorras},
     among many other systems.  Particularly interesting both from the
     theoretical   and  applied  viewpoint   is  the   so-called  {\em
       percolation  threshold}.   Roughly speaking,  it  is the  point
     where the  net is broken  after successive node removal.   In the
     field  of  complex  networks,  percolation thresholds  have  been
     studied considering different  classes of processes, namely, {\em
       intentional   attacks}  \cite{Barabasi:2000}   random  faliures
     \cite{Barabasi:2000},\cite{Random},   \cite{Cohen:2002}   or  the
     so-called  $K$-core descomposition  \cite{Dorogovstevkcore1}.  It
     is worth to note that,  recently, a percolation criteria has been
     derived     even     in     arbitrarily     correlated     graphs
     \cite{Goltsev:2008}.

At  the  theoretical  level,  the  behavior  of  complex  networks  at
criticality  is commonly  studied in  the framework  of  random graph
theory, based on an ensemble formalism \cite{Besesky}, \cite{Bollobas}
close   to  statistical   mechanics   \cite{Park:2004}.   Under   this
framework, some  purely mathematical phenomena, like  the emergence of
the   giant   connected  component   at   the  percolation   threshold
\cite{Erdos:1959} can be understood as a phase transition in the sense
of,  for example,  the transition  from ferromagnetic  to paramagnetic
phase in the  Ising model.  In Landau's theoretical  hallmark to study
phase transitions, one of the main features of a system at criticality
is that some thermodynamical magnitude $m$ (the order parameter of the
system)   displays   a  singularity   in   one   of  its   derivatives
\cite{Stanley:1971}.  Under  mild assumptions,  one can show  that $m$
approaches to  the singularity as a  power law of  a control parameter
$\epsilon$,  i.e.,  $m(\epsilon)\propto \epsilon^{\beta}$.   

Fully embedded  in the mathematical hallmark  briefly described above,
this  theoretical paper  presents a  novel study  on  percolation. The
approach considers  a wide class  of node removal processes  which, in
some  way,  can be  understood  as  the  conceptually inverse  of  the
intentional attacks.  In intentional  attacks, the probability of node
removal  is  defined  according  to  a  direct  relation  between  its
connectivity,  which results  in the  elimination of  nodes displaying
high connectivity, having a specially drammatic effect over scale-free
networks \cite{Barabasi:2000}.  The objective of the present study explores the behavior of
the network when performing an iterative operation of node removal and
where  the probability  for a  node to  be eliminated  has  an inverse
relation to  its connectivity. Specifically,  I study the  behavior at
criticality    of    the     so-called    {\em    nested    subgraphs}
\cite{Nested:2008}.  Nested subgraphs are  a collection of families of
subgraphs of a given graph  whose members can be ordered by inclusion.
We  assume that  these  subgraphs are  obtained  through an  arbitrary
algorithm  whose outcome holds  some probabilistic  requirements.  The
main achievement of the developed  formalism is that the results are
not  linked to  an  specific subgraph,  but  they are  general to  all
processes generating subgraphs satisfying a small set of probabilistic
constraints.   Among others,  we  identify as  nested subgraphs  the
families of  $K$-cores \cite{Dorogovstevkcore1}, \cite{CoresFernholz},
the  $K$-scaffolds  \cite{Shals},  \cite{scaffold}  or  the  subgraphs
obtained   through   random   deletion   of   nodes   \cite{Callaway},
\cite{Random}.   
Furthermore, it can be shown  that  the  degree
distribution of a scale-free network with exponent higher than $2$ is
invariant under nesting operations \cite{Nested:2008}.

The behavior of cluster  sizes at criticality was analitycally studied
from  the first  time  in \cite{Cohen:2000}.   Here  this behavior  is
studied  following   a  methodology  based   on  generating  functions
\cite{Wilf:2004}, introduced  for the  first time in  \cite{Luczak} to
study the  emergence of the  $GCC$.  However, in  this work I  use the
proposal                            made                            in
\cite{Callaway},\cite{Cohen:2002},\cite{Random},\cite{Moore:2000},
also  based  on  the  generating function  formalism,  which  revealed
specially suitable to study  network phenomena from the physical point
of view.  
With  this  mathematical apparatus  I
show that the  probability distribution for the size  of components at
criticality follows a  power-law with universal exponent $\gamma=3/2$,
no matter  the kind  of subgraph is  emerging.  Previous  work derived
this  exponent  for ordinary  percolation  \cite{Random}  and for  the
emergence  of   the  giant  $K$-core   \cite{DorogovtsevKcore2}.   The
relevance of this result comes from the wide variety of processes that
lead to  the emergence  (or disappearance) of  the $GCC$ which  can be
included  in the  category of  nesting  operations.  

To  end with,  we
observe  that the present  paper is  focused on  the emergence  of cluster
sizes  at  criticality  when  an  uncorrelated  nesting  algorithm  is
applied, which implies that the probabilty of removal or survival of a
given node can be expressed as  a function of its connectivity and, in
the extreme cases, of both  its connectivity and a mean field approach
of  the connectivity  of its  first  neighbors.  We  observe that  the
possible  long range  dependencies conditioning  the emergence  of the
$K$-core automatically rules out such  a subgraph from our study, even
a probabilistic  interpretation of the  probability of removal  can be
defined \cite{Nested:2008}.

\section{Uncorrelated Nested Subgraphs}
In this section the main definitions and derivations concerning nested
subgraphs are provided. The presented methodology is based
on  generating functions  \cite{Wilf:2004}, \cite{Callaway},\cite{Random},\cite{Moore:2000}.  In this
framework, the study of the emergence of the $GCC$ resembles the study
of   phase   transitions    under   Landau's   theoretical   framework
\cite{Stanley:1971}.   The  main body  of  this  section is  developed
according to \cite{Nested:2008}.

Formally,  a complex  network is  topologically described  by  a graph
${\cal   G}(V,  \Gamma)$   where  $V$   is  the   set  of   nodes  and
$\Gamma\subseteq V\times V$ the set  of edges connecting nodes of $V$.
If  $p_k$  is the  probability  that a  randomly  chosen  node $e$  is
connected to $k$ other nodes  (noted $d(e)=k$), then the collection of
$p_k$'s  defines a sequence  of real  numbers $\{p_k\}_{k=1}^{\infty}$
(the  so-called degree  distribution) whose  generating  functions are
\cite{Wilf:2004}:
\begin{equation}
g_0(z)=\sum_kp_kz^k;\;\;g_1(z)=\frac{1}{\langle k \rangle}\frac{d}{dz}g_0(z),\nonumber\\
\end{equation}
where 
\begin{eqnarray}
\langle k  \rangle &=&\frac{d}{dz}g_0(z)|_{z=1}=\sum_k^{\infty} kp_k, \nonumber
\end{eqnarray}
is  the  average connectivity  of  ${\cal  G}$.   We assume  that  our
$\{p_k\}^{\infty}_{k=1}$ is, at least, $1$-smooth, i.e., that $\langle
k \rangle <\infty$ \cite{MolloyReed}.

We   will   say   that   $S(A,\Gamma_A)$  is   an   induced   subgraph
\cite{Bollobas:1998} of  ${\cal G}(V, \Gamma)$ if $A  \subseteq V$ and
$\Gamma_A  \subseteq \Gamma$ being  
\begin{equation}
\Gamma_A=\Gamma \bigcap A\times  A.  \nonumber
\end{equation}
A $K$-nested  family of  subgraphs  ${\cal N}$  \cite{Nested:2008} is  a
collection of subgraphs of a  given graph ${\cal G}$ whose members can
be ordered by inclusion\footnote{Notice that such order relation needs
  not to be linear.}:
\begin{equation}
...S_{K+1}({\cal  G})\subseteq  S_K({\cal  G})\subseteq  S_{K-1}({\cal G})...\nonumber
\label{nest}
\end{equation}
Let  the graph  $S_K=S_K(V_{S_K}, E_{S_K})$  be a  member of  a nested
family of subgraphs of a given  graph ${\cal G}$.  For every family of
$K$-nested   subgraphs  we   associate  a   {\em   nesting  function},
$\varphi_K(k)$,  namely the  probability  for a  randomly chosen  node
$e\in V$ with degree $d(e)=k$ to belong to $S_K$:
\begin{equation}
\varphi_K(k)=\mathbf{P}(e\in V_{S_K}|d(e)=k).\nonumber
\end{equation}
Since  $\varphi_K(k)$ is  a  probability,  we can  express  it like  a
function, 
\begin{equation}
\varphi_K(k):U\times \mathbf{N}\rightarrow [0,1],\nonumber
\end{equation}
where $U\subseteq \mathbf{R}$  is a set that depends  on the nature of
the nesting.  We  need our nesting functions to  fulfill the following
conditions:  
\begin{enumerate}
\item
fixed   $K$,  $\varphi_K(k)$  is  a  non-decreasing
function on  $k$,  
\item
fixed  $k$, $\varphi_K(k)$ is  a non-increasing
function on  $K$ and,
\item
  $(\forall K)[(\exists \lambda_{S_K}
\in (0,1])|(\lim_{k \to \infty}\varphi_K(k))= \lambda_{S_K}]$,
where  $\lambda_{S_K}$ is  a scalar  whose  value will  depend on  the
explicit  form of  the nesting  algorithm. 
\end{enumerate}
From the  above properties, we can  conclude that, for  any fixed $K$,
and $(\forall \delta>0)$, $(\exists k^*\in \mathbf{N})$ such that:
\begin{equation}
(\forall k_i,k_j>k^*)(|| \varphi_K(k_i)-\varphi_K(k_j)||<\delta),\nonumber
\end{equation}
and we can conclude that the sequence
\begin{equation}
\{\varphi_K(k)\}_1^{\infty}=\varphi_K(1), \varphi_K(2),..., \varphi_K(i),...\nonumber
\end{equation}
is a Cauchy  sequence.  We observe that a  nesting function takes into
account all the nodes satisfying the imposed conditions: our subgraphs
are maximal under the conditions imposed by the nesting function.

We must  be aware of two  relevant facts: The first  is the underlying
assumption  that,  if  $S_K^{\phi},   S_K^{\varphi}$  are  a  pair  of
subgraphs  of  ${\cal  G}$  -whose associated  nesting  functions  are
$\phi_K$ and $\varphi_K$, respectively- we assume that:
\begin{equation}
(\forall k)(\phi_K(k)>\varphi_K(k))\rightarrow(S_K^{\varphi}   \subseteq S_K^{\phi}).\nonumber
\end{equation}
Secondly, we shall see that, in general:
\begin{equation}
S_{K+1}(S_K({\cal G}))\neq S_{K+1}({\cal G}),
\end{equation}
even in some cases the equality  holds, such as in the $K$-scaffold of
in  the $K$-core  -Although the  latter  cannot be  studied using  the
formalism proposed in this paper.

Let  us define the  generating functions  for an  arbitrary $K$-nested
subgraph with an associated nesting function $\varphi_K(k)$ defined on
a    graph   ${\cal   G}$    with   arbitrary    degree   distribution
$\{p_k\}_{k=1}^{\infty}$.   To be  precise, we  are talking  about the
generating     functions      associated     with     the     sequence
\begin{equation}
\{\varphi_K(k)p_k\}_{k=1}^{\infty} \nonumber
\end{equation}
of real numbers:
\begin{eqnarray}
f_0(z)&=&\sum_{k}^{\infty}p_k\varphi_K(k)z^k \nonumber\\
f_1(z)&=&\frac{1}{\langle k \rangle}\frac{d}{dz}f_0(z)=
\frac{1}{\langle k \rangle}\sum_k^{\infty}kp_k\varphi_K(k)z^{k-1}\nonumber
\end{eqnarray}
Notice  that,   generally,  $f_0(1),  f_1(1)<1$.   For   the  sake  of
completeness, the  section ends  with the assymptotic  expression that
accounts  for  the  degree   distribution  of  the  nested  subgraphs,
$p_{S_K}$. We first notice that  the probability for a {\em surviving}
node displaying connectivity $k$ in ${\cal G}$ to display connectivity
$k'\leq k$ in $S_{K}({\cal G})$ is:
\begin{equation}
\mathbf{P}(k'|k)={k \choose k'}(f_1(1))^{k'}(1-f_1(1))^{k-k'},
\end{equation}
a  relation already  derived in  \cite{Cohen:2000}.  Thus,  the degree
distribution of $S_K({\cal G})$ will be:
\begin{eqnarray}
p_{S_K}(k)&=&\frac{1}{f_0(1)}\sum_{i\geq k}\varphi_K(i){i\choose k}(f_1(1))^k(1-f_1(1))^{i-k}p_i\nonumber\\
&=&\frac{\lambda_{S_K}}{f_0(1)}\sum_{i\geq k}\left[{i\choose k}(f_1(1))^k(1-f_1(1))^{i-k}p_i\right.\nonumber\\
&&-\frac{\lambda_{S_K}}{f_0(1)}\sum_{i\geq k}(\lambda_{S_K}-\varphi_K(i))\times\nonumber\\
&&\times\left.{i\choose k}(f_1(1))^k(1-f_1(1))^{i-k}p_i\right]\nonumber\\
&\approx&\frac{\lambda_{S_K}}{f_0(1)}\sum_{i\geq k}{i\choose k}(f_1(1))^k(1-f_1(1))^{i-k}p_i\nonumber\\
&=&\left.\frac{\lambda_{S_K}}{f_0(1)}\frac{(f_1(1))^k}{k!}\frac{d^k}{dz^k}g_0(z)\right|_{z=1-f_1(1)}.\nonumber
\end{eqnarray}
We  observe  that  the  third  step  is  valid  from  the  fact  that
$\{\varphi_K(k)\}_1^{\infty}$ is a Cauchy sequence.

\section{Behavior at Criticality}
Once the  operation of nesting is accomplished,  the obtained subgraph
can  display many  components  of several  sizes,  including, in  some
cases,  one  component of  {\em  infinite}  size  containing a  finite
fraction of all nodes, the  $GCC$.  Let $\pi_s$ be the probability for
a randomly  chosen node $e$ to  belong to a component  with $s$ nodes.
We observe  that the collection of  $\pi_s$'s form a  sequence of real
numbers   $\{\pi_s\}^{\infty}_{s=1}$   whose   associated   generating
functions are:
\begin{eqnarray}
h_0(z)&=&\sum_{s}^{\infty}\pi_sz^s \label{h0}\\
h_1(z)&=&\frac{1}{\langle s \rangle}\frac{d}{dz}h_0(z)=\frac{1}{\langle s \rangle}\sum_s^{\infty}s\pi_s z^{s-1}\label{h1}
\end{eqnarray}
being  
\begin{equation}
\langle s  \rangle  =\left.\frac{d}{dz}h_0(z)\right|_{z=1},\nonumber 
\end{equation}
the
average size of  components other than the $GCC$.   If $h_0(1)$ is the
probability that a randomly chosen node  $e$ is not in the $GCC$, then
the  probability  for  such a  node  to  belong  to the  $GCC$,  noted
$\pi_{\infty}$, will be
\begin{equation}
\pi_{\infty}=1-h_0(1).\nonumber
\label{piinfty}
\end{equation}
However, this formulation does not  help us to understand the problem.
Following techniques  close to the  ones developed to  study branching
processes,  we can  find  an  alternative form  for  $h_0$ and  $h_1$.
Indeed, it  can be shown  that $h_1$ displays a  Dyson-like recurrence
relation \cite{Moore:2000}, \cite{Callaway}, \cite{Random}:
\begin{eqnarray}
h_1(z)&=&1-f_1(1) +z\frac{p_1}{\langle k\rangle}\varphi_K(1)+ z \frac{2p_2}{\langle k \rangle}\varphi_K (2)h_1(z)
+\nonumber\\&+& z \frac{3p_3}{\langle k \rangle}\varphi_K (3)h^2_1(z)+...\nonumber\\
&=&1-f_1(1)+z\sum\frac{kp_k}{\langle k \rangle}h^{k-1}_1(z)\nonumber\\
&=&1-f_1(1)+zf_1(h_1(z))
\label{H1}
\end{eqnarray}
and  that the generating  function for  the size  of the  component to
which a randomly chosen node belongs to is:
\begin{equation}
h_0(z)=1-f_0(1)+zf_0(h_1(z)).
\label{H0}
\end{equation}
With this formulation, 
\begin{equation}
\pi_{\infty}= f_0(1)-f_0(u),
\label{piinfty}
\end{equation}
where $u$  is the first,  non-trivial solution of  the self-consistent
equation     $u=1-f_1(1)+f_1(u)$    \cite{Callaway},    \cite{Random}.
Furthermore, from the above definition of $h_0$ we can obtain a useful
expression of $\langle s \rangle$:
\begin{eqnarray}
\langle s  \rangle& =&\left.\frac{d}{dz}h_0(z)\right|_{z=1}\nonumber\\
&=&f_0(1)+\left.\frac{d}{dz}f_0(z)\right|_{z=1}\frac{f_1(1)}{1-\left.\frac{d}{dz}f_1(z)\right|_{z=1}}
\label{sing}
\end{eqnarray}
As in  modern theory  of phase transitions,  the main feature  of this
phase  transition   is  the  existence   of  a  singularity   in  some
thermodynamic/statistical magnitude  \cite{Stanley:1971}. In our case,
the phase transition can be identified with the singularity we find in
the component size distribution, $\langle s \rangle$ (eq. \ref{sing}),
at:
\begin{equation}
\left.\frac{d}{dz}f_1(z)\right|_{z=1}=1.
\label{D1}
\end{equation}
Before  the  transition,  $\pi_{\infty}=0$,  being all  components  of
finite  size  and, after  the  transition,  $\pi_{\infty}>0$, and  the
remaining  components display still  finite size.   Specifically, from
eqs.   (\ref{piinfty},   \ref{D1})  \cite{Random},  \cite{MolloyReed},
\cite{Nested:2008} it can be shown that if:
\begin{equation}
\sum_k k(k-2)p_k>\sum_k k(k-1)(1-\varphi_K(k))p_k\nonumber
\label{Percol1}
\end{equation}
then there  exists a  single component of  infinite size  containing a
finite fraction of nodes, i.e., the $GCC$.  The {\em phase transition}
referred also  as the {\em  percolation threshold}, is located  at the
point where:
\begin{equation}
\sum_k k(k-2)p_k=\sum_k k(k-1)(1-\varphi_K(k))p_k.\nonumber
\label{Percol0}
\end{equation}
The critical region  is located near the percolation  threshold (if it
exists), i.e., in the region near (\ref{D1}). To study the behavior of
the cluster  size distribution  near the singularity,  we look  at the
expression   of  $h_0(z)$,   both  depending   on  $f_0$   and  $h_1$.
Nevertheless,  we assume that  $\frac{d}{dz}f_0(z)$ converges  for any
$|z|\leq 1$ (i.e.  $\langle  k \rangle<\infty$ and well defined, being
$\{\varphi_K(k)p_k\}_{k=1}^{\infty}$  at least  $1$-smooth).   Thus we
must  look for  the singularity  in $h_1$.   To study  $h_1$  near the
transition, we define its functional inverse, $h_1^{-1}(\tau)=z$:
\begin{equation}
h_1^{-1}(\tau)=\frac{\tau-1+f_1(1)}{f_1(\tau)}
\label{H-1}
\end{equation}
Note  that, due  to the  fact  that all  the members  of the  sequence
$\{\varphi_K(k)p_k\}_{k=1}^{\infty}$ are not  negative, we can be sure
that the  of zeros  of $f_1$ fall  outside the  statistically relevant
region -and, hence  the poles of $h_1^{-1}$.  Thus  we assume, without
any loss of generality,  that $f_1(z)\neq 0$.  Consistently, we expect
to     find     the     singularity     at     the     point     where
\begin{equation}
\left.\frac{d}{d\tau}h_1^{-1}(\tau)\right|_{\tau=\tau^*}=0.\nonumber 
\end{equation}
Differentiating eq. (\ref{H-1}), we see that:
\begin{equation}
f_1(\tau^*)-(\tau ^*-1+f_1(1))\left.\frac{d}{d\tau}f_1(z)\right|_{z=\tau ^*}=0\nonumber
\end{equation}
As we  argued above, we expect  the phase transition of  the system to
occur at $\frac{d}{dz}f_1(z)|_{z=1}=1$.   Thus, if $\tau^*=1$, all the
terms are  cancelled.  Furthermore, from (\ref{H-1}) we  can see that,
if $\tau^*=1$,  then, $z^*=1$.  Collecting the  above ingredients, and
assuming that the  network is such that $h_1^{-1}$  is analytical near
the singularity of $h_1$, we can perform the power series expansion of
$h_1^{-1}$ about $1$:
\begin{eqnarray}
h_1^{-1}(z)&=& 1+\sum_{i=1}^{\infty}\frac{1}{i!}\frac{d^i}{dz^i}h_1^{-1}(z)|_{z=1}(1-z)^i\nonumber\\
&=&1-\left.\frac{1}{2f_1^2(1)}\frac{d^2}{dz^2}^2f_1(z)\right|_{z=1}(1-z)^2+{\cal O}(1-z)^3\nonumber
\end{eqnarray}
(recall  that that  $\frac{d}{d\tau} h_1^{-1}|_{\tau=1}=0$).  We  can
assume without any loss of generality that:
\begin{equation}
\left.\frac{1}{2f_1^2(1)}\frac{d^2}{dz^2}f_1(z)\right|_{z=1}\neq 0\nonumber.
\end{equation}
Thus, knowing that $h_1^{-1}\left(h_1(z)\right)=z$, we are legitimated
to say that, near $z=1$:
\begin{equation}
z\approx 1-\left.\frac{1}{2f_1^2(1)}
\frac{d^2}{dz^2}^2f_1(z)\right|_{z=1}(1-h_1(z))^2\nonumber
\end{equation}
This enables us to find the exponent $\beta$, indicating the power-law
behavior of $h_1(z)$ near the singularity. Specifically,
\begin{equation}
h_1(z)\approx 1-c\sqrt{1-z}\nonumber
\end{equation}
being  $c$ a constant  depending on  the values  of both  $f_1(1)$ and
$\frac{d^2}{dz^2}f_1(z)|_{z=1}$.     Thus,   near    the   transition,
$h_1(z)\propto  (1-z)^{\beta}$, with $\beta  =1/2$, the  standard mean
field exponent.  We observe that $h_0(z)$ behaves identically near the
singularity.  Indeed, if we are close to $z=1$:
\begin{eqnarray}
h_0(z)\propto (1-z)^{\beta}+{\cal O}(1-z).\nonumber
\end{eqnarray}
However, we did  not end the job, since we are  also interested in the
cluster size  probability distribution $\{\pi_s\}_{s=1}^{\infty}$.  We
attack the  problem by expanding in  power series the  leading term of
$h_0(z)$ when $z$ is close to $1$:
\begin{eqnarray}
h_0(z)\propto \sum_s^{\infty}{\beta \choose s}(-1)^sz^s+{\cal O}(1-z)\nonumber
\end{eqnarray}
Notice that we have an  approximation of $h_0(z)$ in its original form
given in eqs. (\ref{h0},\ref{h1}). Thus, 
\begin{eqnarray}
\pi_s &=&\left.\frac{1}{s!}\frac{d^s}{dz^s}h_0(z)\right|_{z=0}\nonumber\\
&\propto &{\beta \choose s}(-1)^s\nonumber\\
&=&{s-\beta-1 \choose s}\nonumber\\
&=&\frac{\Gamma(s-\beta)}{\Gamma(-\beta)\Gamma(s+1)}\nonumber\\
&\approx&\frac{1}{\Gamma(-\beta)}\left(\frac{s-\beta-1}{e}\right)^{s-\beta-1}\sqrt{2\pi(s-\beta-1)}\nonumber\\
&&\times\left[\left(\frac{s}{e}\right)^{s}\sqrt{2\pi s}\right]^{-1}\nonumber\\
&\approx &\frac{(es)^{-(1+\beta)}}{\Gamma(-\beta)} \nonumber
\end{eqnarray}
where, in this case $\Gamma$, refers to the ordinary Gamma Function and
the last step is obtained  assuming $s\to \infty$ and, hence, applying
Stirling's approach \cite{Abramowitz:1972}.  Since $\beta=1/2$, in the
limit of large $s$:
\begin{eqnarray}
\pi_s\propto s^{-\gamma}; \;\;\;\gamma=1+\beta=\frac{3}{2}.
\end{eqnarray}
It is worth noting that this exponent coincides with the one found for
ordinary  percolation \cite{Random}  and  the emergence  of the  giant
$K$-core\cite{DorogovtsevKcore2}.

\section{Discussion}
In  this  short note  I  demonstrated that  a  wide  variety of  graph
processes  display the  same behavior  at  criticality.  Specifically,
given  any iterative  nesting operation,  we expect  the  cluster size
distribution   to  follow   a  power   law  with   universal  exponent
$\gamma=3/2$ at the critical region where the giant component emerges.
As  pointed out concerning  the $K$-core  in \cite{DorogovtsevKcore2},
the  emerging  components  do  respect the  connectivity  requirements
imposed  by  the nesting  algorithm,  a  feature  that goes  far  from
ordinary   percolation,  where   only   {\em  to   be  connected}   is
required. Beyond  its intrinsic theoretical interest,  the broad class
of mechanisms that can be  described through a nesting algorithm makes
the  universality of  this result  potentially powerful  to understand
natural  phenomena at  criticality where  some kind  of non-correlated
pruning/addition  process is  at  work. Furthermore,  it  is worth  to
emphasize  that the  fact  that the  standard  mean-field exponent  is
obtained does not imply that the derived results can be reduced to the
ordinary  percolation  considering a  random  deletion  of nodes  with
probability $p$. The  reason stems from the fact  that one can compute
the average probability of removal of any deletion process -correlated
or  uncorrelated. Some  of them,  such  as the  case of  intentionated
attacks,  would lead  the analytic  treatment  to failure,  for it  is
important how  the deletion takes place.   Therefore, nested subgraphs
form a  general, well  defined class of  graph processes by  which the
behavior  at criticality  is close  to the  one observed  for ordinary
percolation -which,  as it can be  observed, falls in  the category of
processes studied in this paper.   Further works should study the role
of correlations  in both prunning  algorithms and target  networks, or
the behavior  at criticality of  the nets whose series  expansions are
not analytic, following the unifying philosophy underlying the concept
of nested subgraph.

\section{Acknowledgments}
The author thanks Mari\'an  Bogu\~{n}\'a i Espinal for useful comments
and for  finding mistakes  in the former  manuscript and  an anonymous
reviewer for his/her useful  comments on the manuscript. I acknowledge
Andreea Munteanu, Ricard Sol\'e  and Josep Sardany\'es for the careful
reading of the manuscript.  This  work has been supported by the James
McDonnell Foundation.

\end{document}